\documentclass[showpacs,twocolumn,prl,superscriptaddress]{revtex4}
\usepackage[usenames]{color}
\usepackage{graphicx}
\usepackage{amssymb}
\usepackage{amsmath}
\usepackage{amsfonts}
\usepackage{mathrsfs}

\renewcommand{\epsilon}{\varepsilon}
\newcommand{\figurewidth}{0.38\textwidth}

\begin{document}
\title{Sequence dependence of DNA translocation through a nanopore}

\author{Kaifu Luo }
\altaffiliation[]{
Author to whom the correspondence should be addressed}
\email{luokaifu@yahoo.com}
\affiliation{Laboratory of Physics, Helsinki University of Technology,
P.O. Box 1100, FIN-02015 TKK, Espoo, Finland}
\author{Tapio Ala-Nissila}
%\email{tapio.ala-nissila@tkk.fi}
\affiliation{Laboratory of Physics, Helsinki University of Technology,
P.O. Box 1100, FIN-02015 TKK, Espoo, Finland}
\affiliation{Department of Physics, Box 1843, Brown University, Providence,
Rhode Island 02912-1843, USA}
\author{See-Chen Ying}
\affiliation{Department of Physics, Box 1843, Brown University, Providence, Rhode Island
02912-1843, USA}
\author{Aniket Bhattacharya}
\affiliation{Department of Physics, University of Central Florida, Orlando, Florida
32816-2385, USA}

\date{\today}

\begin{abstract}

We investigate the dynamics of DNA translocation through a nanopore
using 2D Langevin dynamics simulations, focusing on the dependence of
the translocation dynamics on the details of DNA sequences. The DNA
molecules studied in this work are built from two types of bases $A$
and $C$, which has been shown previously to have different
interactions with the pore. We study DNA with repeating blocks
$A_nC_n$ for various values of $n$, and find that the translocation
time depends strongly on the {\em block length} $2n$ as well as  on
the {\em orientation} of which base entering the pore first. Thus,
we demonstrate that the measurement of translocation dynamics of DNA
through nanopore can yield detailed information about its
structure. We have also found that the periodicity of the block
sequences are contained in the periodicity of the residence time of
the individual nucleotides inside the pore.
\end{abstract}


\maketitle
%\section{Introduction}%%%%%%%%%%%%%%%%%%%%%%%%%%%%%%%%

The translocation of biopolymers across membranes is ubiquitous in
biological systems, such as DNA and RNA translocation across nuclear
pores, protein transport through membrane channels, and virus
injection into cells. In a seminal experimental paper, Kasianowicz
\textit{et al.}~\cite{Kasianowicz} demonstrated that an electric
field can drive single-stranded DNA and RNA molecules through the
water-filled $\alpha$-hemolysin channel and that the passage of each
molecule is signaled by a blockade in the channel current, whose
magnitude and duration depend on the structure of the DNA or RNA
molecule. Similar experiments have been done recently using solid
state nanopores with more precisely controlled dimensions. Triggered
by these experiments and potential technological
applications~\cite{Kasianowicz,Meller03}, such as rapid DNA
sequencing, gene therapy and controlled drug delivery, the
translocation of biopolymers through nanopore has become a subject
of intensive 
experimental~\cite{Akeson,Meller00,Meller01,Meller02,Sauer,Storm}
and theoretical
~\cite{Storm,Sung,Muthukumar99,Lubensky,Kafri,Metzler,Chuang,
Milchev,Luo12,Luo3,Luo4,Huopaniemi12,Kotsev,Tsuchiya} studies.
%%%%%%%%%%%%%%%%%
A particular important question is if DNA translocation through a 
nanopore can be used to determine the detailed sequence structure 
of the molecule~\cite{Akeson,Meller00,Meller03,Luo3}.

%%%%%%%%%%%%%%%%%%%%%%
It has been demonstrated in experiments~\cite{Meller00,Meller02} that
translocation through $\alpha$-hemolysin pore can be used to discriminate 
between polydeoxyadenylic acid (poly(dA)) and polydeoxycytidylic acid 
(poly(dC)) molecules of the same chain length.
The translocation time of poly(dA) is found to be longer, and its 
distribution is wider with a longer tail compared with the corresponding 
data for poly(dC). 
The different behavior was attributed to different interactions of the 
nucleotides with the pore, with the base $A$ having a stronger attractive 
interaction with the pore than the base $C$. 
These experimental findings and conclusions were quantitatively supported
by recent Langevin dynamics (LD) simulations~\cite{Luo4} with a model for 
the DNA molecules incorporating different base-pore interactions.

Inspired by the ability to discriminate poly(dA) and poly(dC) with
the same chain length,
%%%%%%%%%%%%%%%%%%%%%% 
Meller \textit{et al.}~\cite{Meller00,Meller02}
have studied different behavior of the translocation time
distribution for the heteroDNA molecules poly(dA$_{50}$dC$_{50}$)
and poly(dAdC)$_{50}$. The result suggests that translocation
through a nanopore can distinguish between DNA polynucleotides of
similar length with compositions that differ only in detailed
sequence structure. In this paper we seek to shed light on this
important question. We adopt a model for the hetero-DNA molecules
with different base-pore interactions and investigate the sequence
dependence of their translocation dynamics using LD simulations.

%%%%%%%%%%%%%%%%%%%%%%%%%%%%%%%%%%%%%%%%%%%%%%%%%%%%%%%%%%%%%%%%%%%%%%%%
%\section{Model and method} \label{chap-model}%%%%%%%%%%%%%%%%%%%%%%%%%
In our model, a single stranded DNA molecule is represented as a
bead-spring chain. A short range repulsive Lennard-Jones (LJ) potential
$U_{LJ}(r)=4\epsilon[{(\frac{\sigma}{r})}^{12}-{(\frac{\sigma}
{r})}^6]+\epsilon$ for $r\le 2^{1/6}\sigma$ and 0 for
$r>2^{1/6}\sigma$ exists between all beads leading to excluded
volume interaction. Here, $\sigma$ is the diameter of a bead,
and $\epsilon$ is the depth of the potential.
Neighboring beads are connected by a Finite Extension Nonlinear Elastic (FENE)
spring with interaction energy
$U_{FENE}(r)=-\frac{1}{2}kR_0^2\ln(1-r^2/R_0^2)$, where $r$ is the distance
between consecutive monomers, $k$ is the spring constant and $R_0$ is the
maximum allowed separation between connected monomers.
We consider a 2D geometry as shown in Fig. 1, where the wall of thickness
$L$ is formed by columns of stationary particles. A pore of length
$L$ and width $W$ in the center of the wall connects the \textit{cis} and the
\textit{trans} sides and a voltage is applied across the pore to drive the
negatively charged DNA through the pore.
Between all bead-wall particle pairs, there exist the same short range
repulsive LJ interaction as described above. The pore-bead interaction
is modeled by a LJ potential with a cutoff of $2.5\sigma$ and
interaction strength $\epsilon_{pA}$ for the base $A$ and
$\epsilon_{pC}$ for the base $C$. Each bead is subjected to
conservative, frictional, and random forces, respectively, leading
to the equation of motion $m{\bf \ddot {r}}_i =-{\bf
\nabla}({U}_{LJ}+{U}_{FENE})+ {\bf F}_{\textrm{ext}}-\xi{\bf
v}_i+{\bf F}_i^R$, where $m$ is the bead's mass, $\xi$ is the
friction coefficient, ${\bf v}_i$ is the bead's velocity, and ${\bf
F}_i^R$ is the random force which satisfies the
fluctuation-dissipation theorem~\cite{Allen}. The external force due
to the applied voltage is represented by ${\bf
F}_{\textrm{ext}}=F\hat{x}$.

The LJ parameters $\epsilon$, $\sigma$ and the bead mass fix the
system energy, length and mass units respectively, leading to the
corresponding time scale $t_{LJ}=(m\sigma^2/\epsilon)^{1/2}$ and
force scale $\epsilon/\sigma$. In our model, each bead corresponds
to a Kuhn length of a single-stranded DNA containing approximately
three nucleotide bases, so the value of $\sigma \sim 1.5$
nm~\cite{Kuhn_length}. The average mass of a base in DNA is about
312 amu, so the bead mass $m \approx 936$ amu. We set
$k_{B}T=1.2\epsilon$, which means the interaction strength
$\epsilon$ to be $3.39 \times 10^{-21}$ J at actual temperature 295
K. This leads to a time scale of 32.1 ps and a force scale of 2.3
pN. The dimensionless parameters in the model are then chosen to be
$R_0=2$, $k=7$, $\xi=0.7$, $L=5$ and $W=3$, and $F=0.5$.
Each base (nucleotide) is estimated to have an effective charge of 
0.094e from ref.~\onlinecite{Sauer}, leading to an effective charge 
of a bead being 0.282e. 
Thus, $F=0.5$ corresponds to a voltage of about 187.9 mV across the pore 
within the range of experimental parameters~\cite{Kasianowicz,Meller00,
Meller01,Meller02,Meller03}. 
The choice of $W=3$ ensures that the average interaction of both
bases $A$ and $C$ with the pore are attractive. The pore-base
interactions $\epsilon_{pA}=3.0$ and $\epsilon_{pC}=1.0$ are chosen
based on comparison of the theoretical results~\cite{Luo4} with the
experimental data ~\cite{Meller00} for the transloation time 
distribution histogram of the homoDNA molecules poly(dC)$_{100}$ and
poly(dA)$_{100}$.
%%%%%%%%%%%%%%
The Langevin equation is integrated in time by a method described by
Ermak and Buckholz~\cite{Ermak} in 2D.
Initially, the first monomer of the chain is placed in the entrance of
the pore, while the remaining monomers are under thermal collisions described
by the Langevin thermostat to obtain an equilibrium configuration.

\begin{figure}
  \includegraphics*[width=\figurewidth]{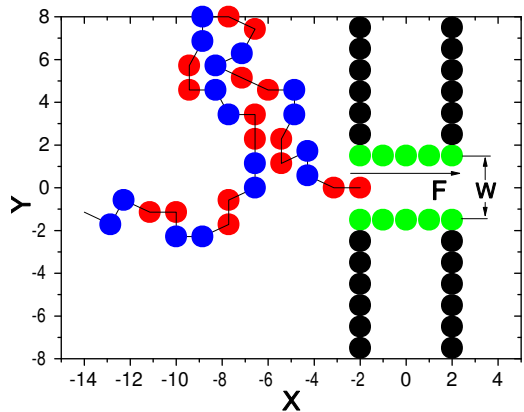}
\caption{ A schematic representation of the system. The pore length
$L=5$ and the pore width $W=3$.
        }
 \label{Fig1}
\end{figure}

%%%%%%%%%%%%%%%%%%%%%%%%%%%%%%%%%%%%%%%%%%%%%%%%%%%%%
%\section{Results and Discussions}

%\subsection{Sequence dependence on the block length for symmetric blocks}
We consider the sequence dependent translocation results for DNA of
chain length $N=128$ with the symmetric blocks $A_nC_n$ having block
length $M=2n$, with minimum value of $n=1$ for poly(dAdC)$_{64}$ and
maximum value of $n=N/2$ for poly(dA$_{64}$dC$_{64}$). Fig. 2 shows
the translocation time $\tau$ as a function of the block length.
%%%%%%%%%%%%%%%%%%%%%%%
The translocation time is obtained as the time interval between the
entrance of the first bead into the pore and the exit of the last
bead \cite{comment1}. Typically, we average our data over 2000
independent runs.
%%%%%%%%%%%%%%%%%%%%%%%
The horizontal dashed, dotted and dash-dotted
lines correspond to $\tau_A$, $(\tau_A+\tau_C)/2$ and $\tau_C$,
respectively. Here, $\tau_A$ and $\tau_C$ are the translocation
times for poly(dA)$_{128}$ and poly(dC)$_{128}$, respectively.
%%%%%%%
For $M < 8$, $\tau$ is close to $\tau_C$
and much lower than $(\tau_A+\tau_C)/2$, with very weak dependence
on the block length and the orientation of which monomer enters the
pore first.

\begin{figure}
  \includegraphics*[width=\figurewidth]{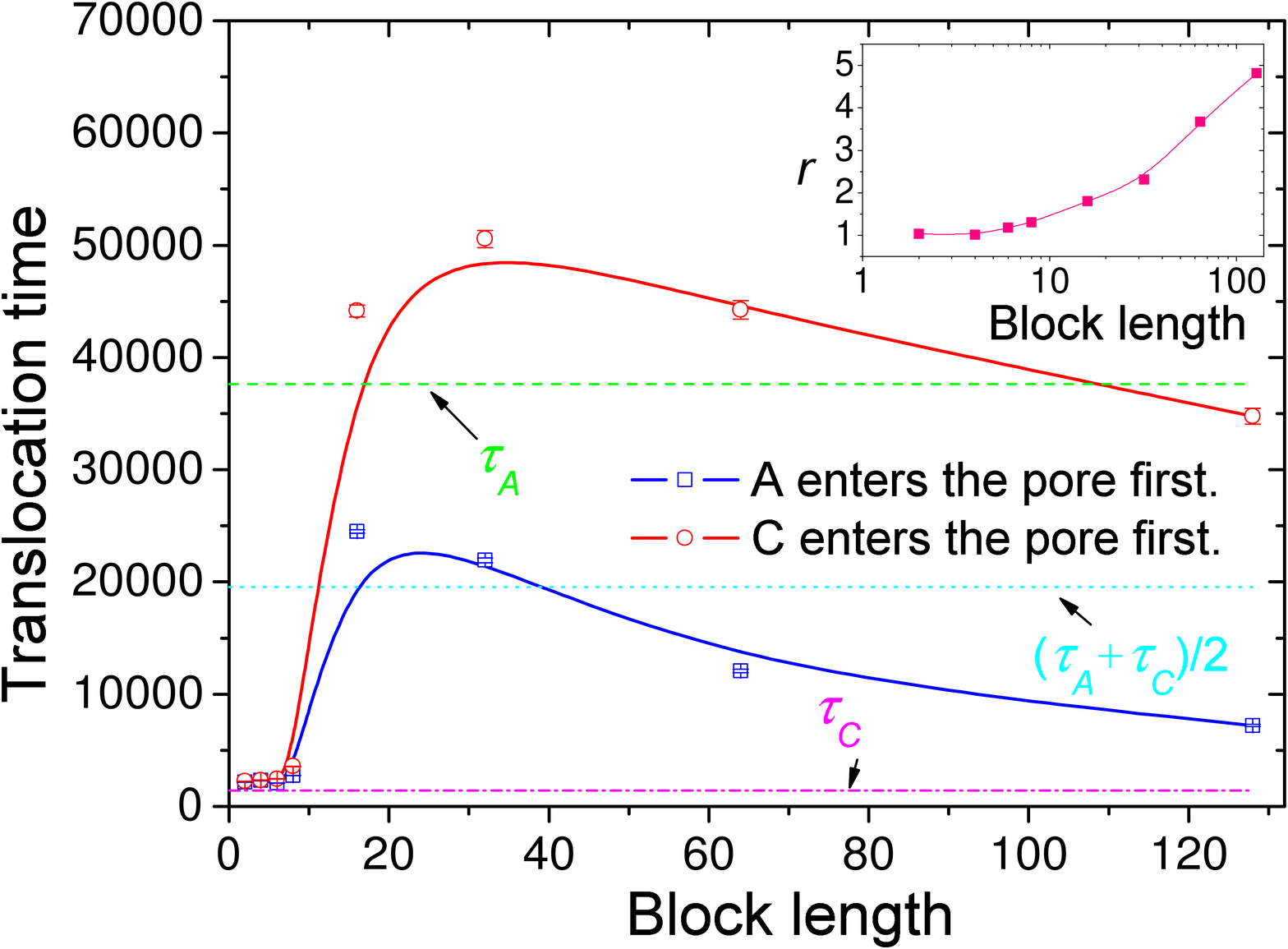}
\caption{Translocation time as a function of the block length for
multi-block DNA with symmetric repeat units $A_nC_n$. Here the block
length $M=2n$, $\epsilon_{pA}=3.0$, $\epsilon_{pC}=1.0$, $F=0.5$,
and the chain length $N=128$.
The insert shows $r$ as a function of the block length for multi-block
DNA with symmetric repeat units $A_nC_n$.
Here, we define $r$ as the ratio of translocation times for the base
$C$ entering the pore first and the base $A$ entering the pore first.
}
 \label{Fig2}
\end{figure}

%%%%
However, $\tau$ begins to increase rapidly with $M$ for $M \ge 8$,
approaches a maximum between $M=16$ and 32, and
finally decreases slowly with increasing $M$. In addition, $\tau$
also depends strongly on the polymer orientation. It is always much
longer for the base $A$ entering the pore first than the other
orientation. Quantitatively, we define $r$ as the ratio of
translocation times for the base $C$ entering the pore first and the
base $A$ entering the pore first. The insert of Fig. 2 shows $r$ as
a function of the block length. For $M \le 4$, $r=1$, but for longer
$M$, $r$ increases exponentially with $M$. For poly(dA$_{64}$dC$_{64}$), 
$r \approx 5$~\cite{comment3}. 

Qualitatively, the large block length results can be understood by
examining the extreme case of the largest block $M=128$. When
the $C_{64}$ block is translocated first through the pore,
subsequent forward motion is energetically unfavored because of the
strong attraction between the pore and the base $A$. As a result,
frequent backward transitions occur which slows down the overall
translocation process. In the opposite orientation when $A_{64}$
block goes through the pore first, the difference of probabilities
between the forward and the backward steps of the remaining motion
is smaller, leading to a value $r>1$. Similar behaviors have been
recently analyzed by Kotsev and Kolomeisky~\cite{Kotsev} where the
translocation of polymers consisting of a double-stranded block and
a single-stranded block is considered, and by Tsuchiya and
Matsuyama~\cite{Tsuchiya} where they studied the translocation of an
amphiphilic polymer.

For the base $C$ entering the pore first, $\tau \gg (\tau_A+\tau_C)/2$
for $M \ge 16$.
It is a surprise that $\tau > \tau_A$ for $16 \le M \le 64$.
For poly(dA), the frequency of backward and forward motion is much slower than
that for poly(dC). Incorporating the base $C$ with a suitable block length
into poly(dA) will increase the frequency of backward and forward motion
when the $C$ block is in the pore.
As a result, the translocation time is larger than $\tau_A$.
For the base $A$ entering the pore first, $\tau > \tau_C$ and for
$16 \le M \le 32$, $\tau > (\tau_A+\tau_C)/2$.

%Experimentally, Meller \textit{et al.}~\cite{Meller00} have
%investigated the translocation of (poly(dA)$_{100}$),
%(poly(dC)$_{100}$), poly(dA$_{50}$dC$_{50}$) and poly(dAdC)$_{50}$
%DNA molecules through the $\alpha$-hemolysin channel.
%For the homoDNAs, the translocation time of poly(dA)$_{100}$ is
%longer with a wider distribution compared with poly(dC)$_{100}$.
%This difference was attributed to stronger attractive interaction
%of poly(dA) with the pore~\cite{Meller02,Meller03}.
%We have recently confirmed this through explicit numerical studies
%using the same bead-spring model outlined earlier~\cite{Luo4}.
Meller \textit{et al.}~\cite{Meller00} have studied blockade signals
for the heteroDNAs poly(dA$_{50}$dC$_{50}$) and poly(dAdC)$_{50}$.
The translocation events are organized into two well-localized
groups with different blockage currents, the origin of which is yet
uncertain \cite{comment2}. Direct comparison with the present study
is further complicated by the fact that poly(dA) molecules have a
higher tendency to form single-stranded base-stacked helices as
compared with poly(dC)~\cite{Meller02,Meller03}, although the
base-pore interaction effect is still expected to be dominant. The
group 2 data at low temperatures show that the translocation time
for poly(dA$_{50}$dC$_{50}$) is longer than that for
poly(dAdC)$_{50}$, in agreement with our finding here that larger
block length of repeat unit leads to a greater translocation time,
as shown in Fig. 2. It would be desirable to have future
experimental tests for the orientation dependence of the
translocation for these heteropolymers as well. 
%\subsection{Histogram of translocation times}
We have also studied
the histograms of translocation time for the hetero-DNAs
poly(dA$_{n}$dC$_{n}$). For short block length $M=2n$, the
histograms depend only weakly on the orientation, shown in Fig.
3(a), and the behavior is close to that of poly(dC). However, for
longer block lengths, the histogram deviates markedly from a
Gaussian with a long exponential tail as shown in Fig. 3(b). This
behavior is in agreement with the experimental
observation~\cite{Meller00}. There is also a strong orientational
dependence, with the histogram for $C$ entering first shifted to
longer translocation times.

\begin{figure}
  \includegraphics*[width=\figurewidth]{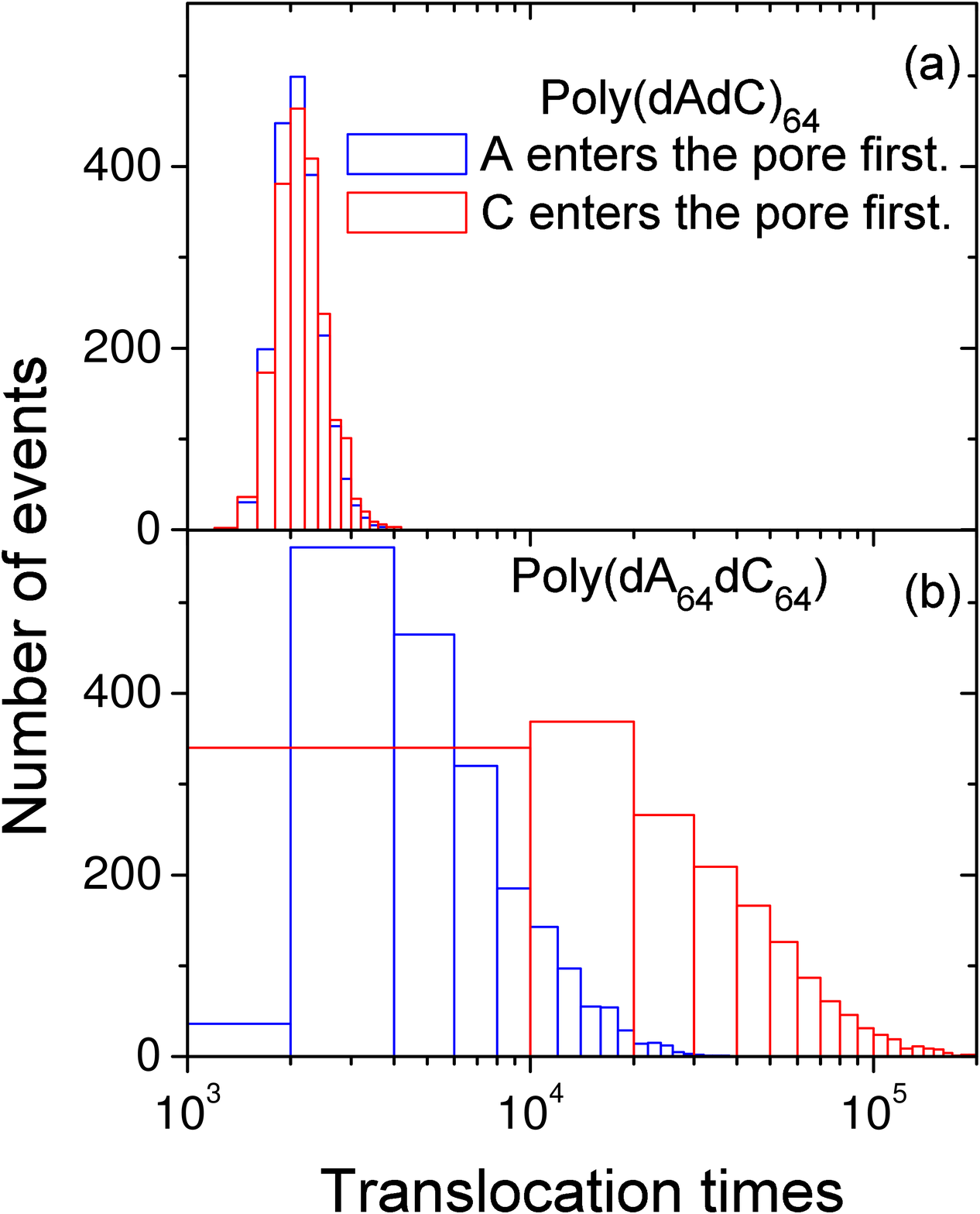}
\caption{Histogram of the translocation times for
(a) poly(dAdC)$_{64}$ and (b) poly(dA$_{64}$dC$_{64}$)
under $F=0.5$.}
 \label{Fig3}
\end{figure}

%\subsection{Waiting time}

\begin{figure}
  \includegraphics*[width=\figurewidth]{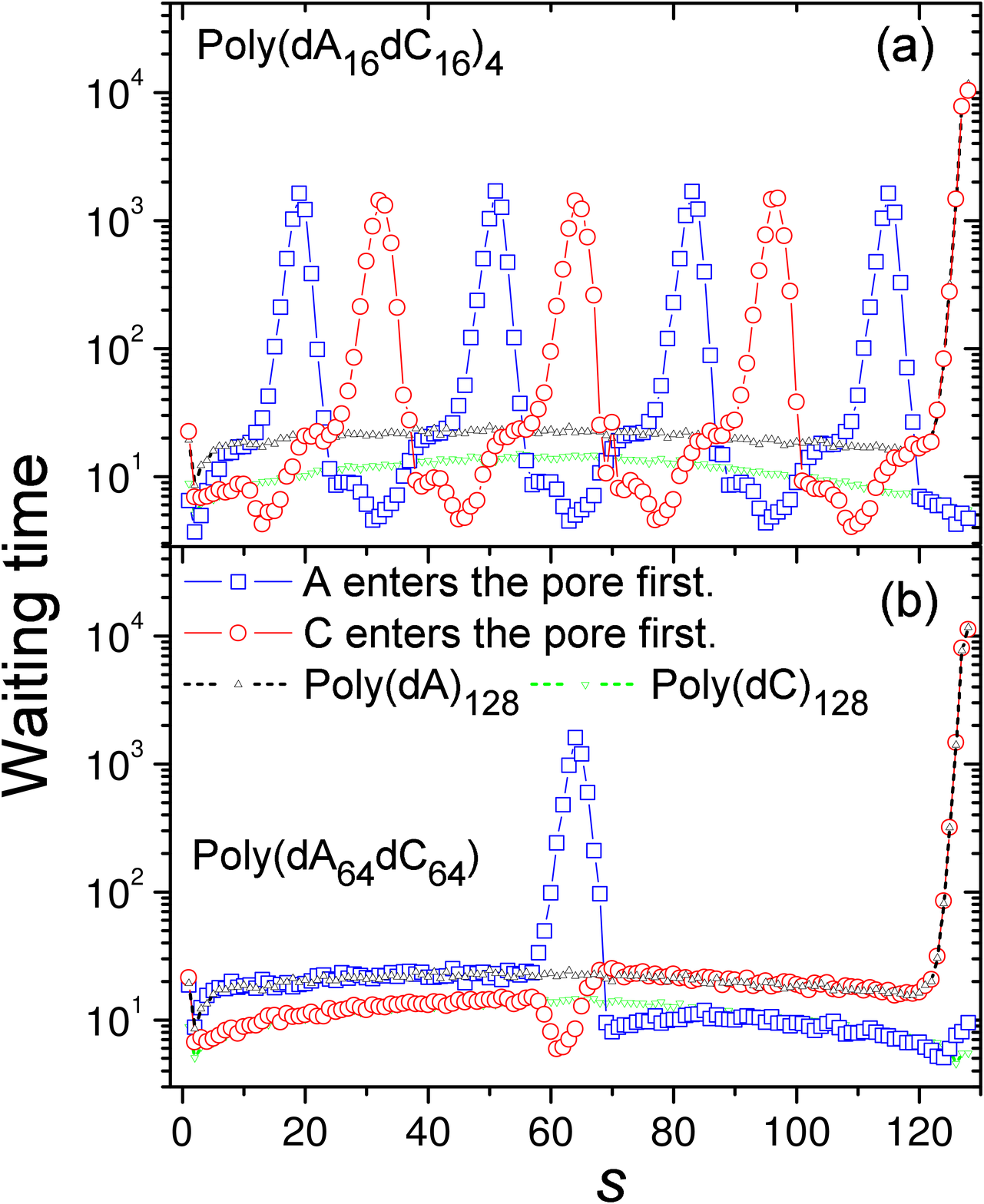}
\caption{Waiting times for (a) poly(dA$_{16}$dC$_{16}$)$_4$ and
(b) poly(dA$_{64}$dC$_{64}$) under $F=0.5$.}
 \label{Fig4}
\end{figure}

We have also investigated the distribution for waiting (residence)
time of base $s$ defined as the time between the events that the
base $s$ and the base $s+1$ exit the pore.
We find  that the residence times for the ordered DNA with repeat
units $A_{n}C_{n}$ (for $n > 2$) exhibit ``fringes'' reminiscent of
optical interference pattern, as shown in Fig. 4.
The number of peaks is exactly equal to $N/2n$. The
periodicity of the waiting time not only depends on the block
length but also the orientational property of the chain as well. For
the sequence $A_nC_n$ with $A$ entering the pore first ($\epsilon_A
> \epsilon_C$), the residence time is symmetric with respect to the
center of the chain containing exactly $N/4n$ maxima on either side,
whereas with $C$ entering the pore first has $(N/2n -1)$ maxima with
the last maximum ending with the largest waiting time.

The sequence dependence of the waiting time distribution yields a
better understanding for the sequence dependence of $\tau$ shown in
Fig. 2. The translocation time can be written as $\tau \sim
\tau_1+\tau_2+\tau_3$, where $\tau_1$, $\tau_2$ and $\tau_3$
correspond to initial filling of the pore, transfer of the base from
the \textit{cis} side to the \textit{trans} side, and finally the
emptying of the pore, respectively. For the present case, $\tau_1
<<\tau_2,\;\tau_3$. For $M \le 8$, $\tau_2$ dominates and it has no
strong dependence on the detailed sequence or the orientation of the
chain.
%%%%%%%%%%%%%%%%%%%
When the base $C$ enters the pore first, $\tau_3$ increases rapidly
with increasing $M$ for $8 \le M \le 16$, and then saturates to a
constant value for $32 \le M \le 128$.
On the other hand, $\tau_2$ is related to both the number of the
``fringes'' and the corresponding maximum time. With increasing $M$,
the former decreases and the latter increases. The interplay of all
these factors leads to a maximum for $\tau$ as a function of the block
length $M$. Similar consideration applies for the base $A$ entering
the pore first except that here $\tau_2$ dominates over $\tau_3$.
%%%%%%%%%%%%%%%%%%%%%%

%\section{Conclusions} \label{chap-conclusions}

To summarize, we have demonstrated that sequences of a driven DNA
can be identified from its translocation specific characteristics
driven through a nanopore that has different affinity for each base.
Simulation studies based on this \textit{attractive nanopore} model
are in accord with the existing experimental data.
A stronger attraction for the polynucleotide $A$ inside the nanopore
leads to a much longer translocation time for (polydA)$_{100}$
as compared to (polydC)$_{100}$.
Further analysis explains the shape of the histogram of the first passage
time, provides an understanding of how translocation time depends on
a specific sequence, and explains the experimental data of longer
translocation time for (polydA)$_{50}$(polydC)$_{50}$ compared to
(polydAdC)$_{50}$. Our simulation studies also reveal a novel
phenomenon that the information for the periodicity of the block
sequences is contained in the periodicity of the residence time
of the individual nucleotides.

\begin{acknowledgments}
This work has been supported in part by The Academy of Finland
through its Center of Excellence (COMP) and TransPoly Consortium grants.
\end{acknowledgments}


\begin{thebibliography}{8}
%%%%%%%%%%%%%%%%%%%%%%%%
\bibitem{Kasianowicz} J. J. Kasianowicz, E. Brandin, D. Branton and D. W. Deamer, \textit{Proc. Natl. Acad. Sci. U.S.A.} {\bf 93}, 13770 (1996).
\bibitem{Meller03} A. Meller, \textit{J. Phys.: Condens. Matter} {\bf 15}, R581 (2003).
%%%%%%%%%%%%%%%Experinemts%%%%%%%%%%%%%%%%%%%%%%%
\bibitem{Akeson} M. Akeson, D. Branton, J. J. Kasianowicz, E. Brandin, and D. W. Deamer, \textit{Biophys. J.} {\bf 77}, 3227 (1999).
\bibitem{Meller00} A. Meller, L. Nivon, E. Brandin, J. A. Golovchenko, and D. Branton, \textit{Proc. Natl. Acad. Sci. U.S.A.} {\bf 97}, 1079 (2000).
\bibitem{Meller02} A. Meller and D. Branton, \textit{Electrophoresis} {\bf 23}, 2583 (2002).
\bibitem{Meller01} A. Meller, L. Nivon, and D. Branton, \textit{Phys. Rev. Lett.} {\bf 86}, 3435 (2001).
%%\bibitem{Henrickson} S. E. Henrickson, M. Misakian, B. Robertson, and J. J. Kasianowicz, \textit{Phys. Rev. Lett.} {\bf 85}, 3057 (2000).
\bibitem{Sauer} A. F. Sauer-Budge, J. A. Nyamwanda, D. K. Lubensky, and D. Branton, \textit{Phys. Rev. Lett.} {\bf 90}, 238101 (2003);
                J. Mathe, H. Visram, V. Viasnoff, Y. Rabin, and M. Meller, \textit{Biophys. J.} {\bf 87}, 3205 (2003).
%%\bibitem{Li} J. L. Li, D. Stein, C. McMullan, D. Branton, M. J. Aziz, and J. A. Golovchenko, \textit{Nature} (London) {\bf 412}, 166 (2001);
%%          J. L. Li, M. Gershow, D. Stein, E. Brandin, and J. A. Golovchenko, \textit{Nat. Mater.} {\bf 2}, 611 (2003).
%%\bibitem{Chen} A. J. Storm, J. H. Chen, X. S. Ling, H. W. Zandbergen, and C. Dekker, \textit{Nat. Mater.} {\bf 2}, 537 (2003).
%\bibitem{Krasilnikov} O. V. Krasilnikov, C. G. Rodrigues, and S. M. Bezruko, \textit{Phys. Rev. Lett.} {\bf 97}, 018301 (2006).
\bibitem{Storm} A. J. Storm, C. Storm, J. Chen, H. Zandbergen, J. -F. Joanny and C. Dekker, \textit{Nano Lett.} {\bf 5}, 1193 (2005).
%%%%%%%%%%%%%%%Theory%%%%%%%%%%%%%%%%%%%%%%%%%%%
%\bibitem{Simon} S. M. Simon, C. S. Peskin, and G. F. Oster, \textit{Proc. Natl. Acad. Sci. U.S.A.} {\bf 89}, 3770 (1992).
\bibitem{Sung}   W. Sung and P. J. Park, \textit{Phys. Rev. Lett.} {\bf 77}, 783 (1996).
%\bibitem{Park} P. J. Park and W. Sung, \textit{J. Chem. Phys.}  {\bf 108}, 3013 (1998).
%\bibitem{diMarzio} E. A. diMarzio and A. L. Mandell, \textit{J. Chem. Phys.}  {\bf 107}, 5510 (1997).
\bibitem{Muthukumar99}   M. Muthukumar, \textit{J. Chem. Phys.} {\bf 111}, 10371 (1999).
%\bibitem{MuthuKumar03} M. Muthukumar, \textit{J. Chem. Phys.}  {\bf 118}, 5174 (2003).
\bibitem{Lubensky} D. K. Lubensky and D. R. Nelson, \textit{Biophys. J.} {\bf 77}, 1824 (1999).
\bibitem{Kafri} Y. Kafri, D. K. Lubensky, and D. R. Nelson, \textit{Biophys. J.} {\bf 86}, 3733 (2004).
%\bibitem{Slonkina} E. Slonkina and A. B. Kolomeisky, \textit{J. Chem. Phys.}  {\bf 118}, 7112 (2003).
%\bibitem{Ambj} T. Ambjornsson, S. P. Apell, Z. Konkoli, E. A. DiMarzio, and J. J. Kasianowicz, \textit{J. Chem. Phys.}  {\bf 117}, 4063 (2002).
\bibitem{Metzler} R. Metzler and J. Klafter, \textit{Biophys. J.} {\bf 85}, 2776 (2003);
       T. Ambjornsson, M. A. Lomholt, and R. Metzler, \textit{J. Phys.: Condens. Matter} {\bf 17}, S3945 (2005).
%\bibitem{Ambj2} T. Ambjornsson and R. Metzler, \textit{Phys. Biol.} {\bf 1}, 19 (2004).
%\bibitem{Baumg} A. Baumgartner and J. Skolnick, \textit{Phys. Rev. Lett.} {\bf 74}, 2142 (1995).
%\bibitem{Chuang} J. Chuang, Y. Kantor and M. Kardar, \textit{Phys. Rev. E} {\bf 65}, 011802 (2002).
%\bibitem{Kantor} Y. Kantor and M. Kardar,  \textit{Phys. Rev. E} {\bf 69}, 021806 (2004).
\bibitem{Chuang} J. Chuang, Y. Kantor and M. Kardar, \textit{Phys. Rev. E} {\bf 65}, 011802 (2002); 
                 Y. Kantor and M. Kardar,  \textit{Phys. Rev. E} {\bf 69}, 021806 (2004).
\bibitem{Milchev} A. Milchev, K. Binder, and A. Bhattacharya, \textit{J. Chem. Phys.} {\bf 121}, 6042 (2004).
%\bibitem{Luo1} K. F. Luo, T. Ala-Nissila, and S. C. Ying, \textit{J. Chem. Phys.} {\bf 124}, 034714 (2006).
%\bibitem{Luo2} K. F. Luo, I. Huopaniemi, T. Ala-Nissila, and S. C. Ying, \textit{J. Chem. Phys.} {\bf 124}, 114704 (2006).
\bibitem{Luo12} K. F. Luo, T. Ala-Nissila, and S. C. Ying, \textit{J. Chem. Phys.} {\bf 124}, 034714 (2006); 
                K. F. Luo, I. Huopaniemi, T. Ala-Nissila, and S. C. Ying, \textit{J. Chem. Phys.} {\bf 124}, 114704 (2006).
\bibitem{Huopaniemi12} I. Huopaniemi, K. F. Luo, T. Ala-Nissila, and S. C. Ying, \textit{J. Chem. Phys.} {\bf 125}, 124901 (2006); \textit{Phys. Rev. E} {\bf 75}, 061912 (2007).
%\bibitem{Huopaniemi2} I. Huopaniemi, K. F. Luo, T. Ala-Nissila, and S. C. Ying, \textit{Phys. Rev. E} {\bf 75}, 061912 (2007).
\bibitem{Luo3} K. F. Luo, T. Ala-Nissila, S. C. Ying, and A. Bhattacharya, \textit{J. Chem. Phys.} {\bf 126}, 145101 (2007).
\bibitem{Luo4} K. F. Luo, T. Ala-Nissila, S. C. Ying, and A. Bhattacharya, \textit{Phys. Rev. Lett.} {\bf 99}, 148102 (2007).
%%%%%%%%%%%%
%\bibitem{Chern} S.-S. Chern, A. E. Cardenas, and R. D. Coalson, \textit{J. Chem. Phys.} {\bf 115}, 7772 (2001).
%\bibitem{Loebl} H. C. Loebl, R. Randel, S. P. Goodwin, and C. C. Matthai, \textit{Phys. Rev. E} {\bf 67}, 041913 (2003).
%\bibitem{Randel} R. Randel, H. C. Loebl, and C. C. Matthai, \textit{Macromol. Theory Simul.} {\bf 13}, 387 (2004).
%\bibitem{Lansac} Y. Lansac, P. K. Maiti, and M. A. Glaser, \textit{Polymer} {\bf 45}, 3099 (2004).
%\bibitem{Kong} C. Y. Kong and M. Muthukumar, \textit{Electrophoresis} {\bf 23}, 2697 (2002).
%\bibitem{Farkas} Z. Farkas, I. Derenyi, and T. Vicsek, \textit{J. Phys.: Condens. Matter} {\bf 15}, S1767 (2003).
%%\bibitem{Tian} P. Tian and G. D. Smith, \textit{J. Chem. Phys.}  {\bf 119}, 11475 (2003).
%\bibitem{Zandi} R. Zandi, D. Reguera, J. Rudnick, and W. M. Gelbart, \textit{Proc. Natl. Acad. Sci. U.S.A.} {\bf 100}, 8649 (2003).
\bibitem{Kotsev} S. Kotsev and A. B. Kolomeisky, \textit{J. Chem. Phys.}  {\bf 125}, 084906 (2006).
\bibitem{Tsuchiya} S. Tsuchiya and A. Matsuyama, \textit{Phys. Rev. E} {\bf 76}, 011801 (2007).
%%%%%%%%%%%%%%%%%%%%%%%%%%%%%%%%%%%%%%%%%%%%%%%%%%%%%%%%%%%
%\bibitem{de Gennes} P. G. de Gennes, \textit{Scaling Concepts in Polymer Physics} (Cornell University Press, Ithaca, NY, 1979).
%\bibitem{Doi} M. Doi, and S. F. Edwards, \textit{The Theory of Polymer Dynamics} (Clarendon, Oxford, 1986).
%%%%%%%%%%%%%%%%%%%%%%%%%%%%%%%%%%%%%%%%%%%%%%%%%%%%%%%%%%%%%%%%%%%%%%%%
\bibitem{Allen}M. P. Allen, D.J. Tildesley, {\it Computer Simulation of Liquids} (Oxford University Press, 1987).
\bibitem{Kuhn_length}S. B. Smith, Y. Cui and C. Bustamante, \textit{Science} {\bf 271}, 795 (1996);
M. C. Murphy, I. Rasnik, W. Cheng, T. M. Loman, and T. Ha, \textit{Biophys. J.} {\bf 86} 2530 (2004).

%\bibitem{Panja} J. K. Wolterink, G. T. Barkema, and D. Panja, \textit{Phys. Rev. Lett.} {\bf 96}, 208301 (2006).
%\bibitem{Holm} H. J. Limbach and C. Holm, \textit{Computer Physics Communications} {\bf 147}, 321 (2002).
\bibitem{Ermak} D. L. Ermak and H. Buckholz \textit{J. Comput. Phys.} {\bf 35}, 169 (1980).

\bibitem{comment1} It is worthwhile to note that the translocation time depends strongly on the
pore length $L$. We have checked that $L \sim 10$ nm produces average
translocation times $\tau \sim $100 $\mu$s in accord with Meller's experiments
\cite{Meller00} for $L \sim 10$ nm. For computational efficiency we present results for $L=5$ here.

\bibitem{comment2} We interpret that the 2nd peak in Meller's experiment \cite{Meller00} corresponds to
pure translocation while the peak 1 may have an unwanted component of a DNA residing at the pore to
block the channel current and finally exiting from the same side it came from.

\bibitem{comment3} We checked $r$ as a function of $\epsilon_{pA}$ for $F=0.5$ and $\epsilon_{pC}=1$ and 
found that the dependence of $\tau$ on chain orientation starts to vanish below 
$\epsilon_{pA}/\epsilon_{pC} \approx 2$. 

\end{thebibliography}
\end{document}